\documentclass[10pt,letterpaper,fleqn]{article}
\usepackage{amsmath}
\usepackage{amsfonts}
\usepackage{fancyhdr}
\usepackage{graphicx}
\usepackage[round]{natbib}
\usepackage{setspace} 
\usepackage{type1cm}
\usepackage{upref}
\usepackage[T1]{fontenc}
\usepackage{times}

\makeatletter
\long\def\@makecaption#1#2{%
  \vskip\abovecaptionskip
  \sbox\@tempboxa{#1 #2}%
  \ifdim \wd\@tempboxa >\hsize
    #1\quad #2\par      
  \else
    \global \@minipagefalse
    \hb@xt@\hsize{\hfil\box\@tempboxa\hfil}%
  \fi
  \vskip\belowcaptionskip}
\def\caption{%
   \ifx\@captype\@undefined
     \@latex@error{\noexpand\caption outside float}\@ehd
     \expandafter\@gobble
   \else
     \refstepcounter\@captype
     \expandafter\@firstofone
   \fi
   {\@dblarg{\@caption\@captype}}%
}
\long\def\@caption#1[#2]#3{%
  \par
  \addcontentsline{\csname ext@#1\endcsname}{#1}%
    {\protect\numberline{\csname the#1\endcsname}{\ignorespaces #2}}%
  \begingroup
    \@parboxrestore
    \if@minipage
      \@setminipage
    \fi
    \normalsize
    \@makecaption{\csname fnum@#1\endcsname}{\ignorespaces #3}\par
  \endgroup}
\makeatother

\def\eqref#1{Eq.~\ref{#1}}

\title{\Large \bf 
Wave Duration/Persistence Statistics, Recording Interval, and Fractal Dimension
      }

\author{\normalsize\em
\def\footnotemark{}
Alastair D. Jenkins\thanks{\hangindent=1em 
\hspace*{-1.8em}The final version of the paper was published in the
International Journal of Offshore and Polar Engineering, vol.~12, no.~2,
pp.~109--113 (2002).  The author is now at the Bjerknes Centre for Climate
Research, University of Bergen, Geophysical Institute, Allégaten 70,
5007 Bergen, Norway, E-mail jenkins@gfi.uib.no.  
}
\\[0ex]
\small\rm Norwegian Meteorological Institute\\
\small\rm Bergen, Norway
}

\date{}

\begin{document}
\maketitle
\bibliographystyle{./abbrvnatIJOPE}

\small 

\begin{abstract}
{\bf
The statistics of sea state
duration (persistence) have been found to be 
dependent upon the recording interval 
$\boldsymbol{\Delta t}$.
Such behavior can be explained as a consequence of the fact that the graph of a
time series of an environmental parameter such as the significant wave height
has an irregular, ``fractal'' geometry.  The mean duration, 
$\boldsymbol{\overline\tau}$ can  
have a power-law dependence on
 $\boldsymbol{\Delta t}$ as $\boldsymbol{\Delta t \to 0}$, with an exponent
equal to the fractal dimension of the level sets of the time series graph.
This recording interval dependence means that the mean duration is not a well
defined quantity to use for marine operational purposes.  A more practical
quantity may be the ``useful mean duration'',
$\boldsymbol{\overline\tau^u}$, estimated from the formula
$\boldsymbol{(\sum\tau_i^2)/(\sum\tau_i)}$, where each interval 
$\boldsymbol{[t_i,t_i+\tau_i]}$ satisfying the
appropriate criterion is weighted by its duration. 
These results are illustrated using wave data from the Frigg gas
field in the North Sea.
}
\end{abstract}


\section*{\rm\normalsize KEY WORDS} 
Duration statistics; mean duration; offshore operations; wave height
persistence; fractal dimension.

\section*{\rm\normalsize INTRODUCTION}

The duration or persistence statistics of sea state and other environmental
parameters are important for purposes such as marine engineering operations, in
which, for example, useful work can only be performed if the significant wave
height $h$ is less than a particular value $h_0$.  Over recent decades a number
of observational and theoretical studies have been made, in order to relate the
statistical behavior of the duration of various sea state criteria to, for
example, the probability distribution of the wave height, its seasonal
variation, and other parameters  
\cite[e.g.][]{%
houmbOG:vikI:poac-1975-241,%
grahamC:ce-1982-303,%
Mathiesen:ce-1994-167,%
tsekosC:anastasiou:aor-1996-243,%
soukissianT:theochari:isope-2001-33%
}.   In this paper we shall consider the collection of time intervals,
in which $h(t)<h_0$: the intervals having durations $\tau_i$, where
$i=1,2,\ldots,N$.
The {\em mean duration}, $\overline\tau$, is calculated using the formula
\begin{equation}
\overline\tau = (1/N)\sum_i\tau_i.
\end{equation}
We can, of course, choose
other criteria, for example $h\geq h_1$, $h_0\leq h<h_1$, etc.

If $h(t)$ is differentiable with respect to $t$, the mean duration
is related to the probability distribution of $h$ and $dh/dt$ by
the Rice--Kac formula \cite[e.g.][]{%
Rice:MARN-1944,%
Kac:bams-1943-314,%
Mathiesen:ce-1994-167%
}:
\begin{equation}
\overline\tau(h_0) = {2 F_h(h_0) \over f_h(h_0)\, E\left[\left| dh/dt \right|
\,\big|\, h=h_0\right]}, \label{eq-rice}
\end{equation}
where $F_h$ is the cumulative distribution function of~$h$, i.e., $F_h(h_0)$ 
is the
probability that $h\leq h_0$, 
$f_h(h_0) = dF_h(h_0)/dh_0$ is the probability
density function of~$h$, and $E\left[\left| dh/dt \right|
\,\big|\, h=h_0\right]$ is the expectation (mean value) of the absolute value of
$dh/dt$, given that $h=h_0$.

However, time series of observed
environmental parameters often have a noisy, irregular appearance, so that
measurements recorded at frequent intervals show considerable structure which
does not appear if the measurements are recorded less frequently. In such a
case, the time derivative is not well defined, so the Rice--Kac formula of 
Eq.~\ref{eq-rice} cannot be applied.
Although this may partly be due to the effect of errors in
the measurement or of sampling variability, the phenomenon has been recognized
as a manifestation of {\em fractal\/} 
behavior, shared with such phenomena as the
irregularity of coastlines, the surfaces of snowflakes, and Brownian motion
\citep{mandelbrotBB:fgn-1983}.  It is a general property of such fractal
objects or curves that their irregular shape remains even when you examine them
at finer and finer scales.  A fractal curve, such as a coastline, does not
have a well defined length, and if you measure it with ``rulers'' of finer and
finer length, the total length will increase without bound.  In order to
measure such irregular objects, it is helpful to generalize the concept of
``dimension'' to non-integer values, and this was done by Hausdorff and
Besicovitch in the early part of the twentieth century
{\frenchspacing\cite[e.g.][]{hausdorff:ma-1919-157}}.  A fractal curve,
since it has an unbounded length, but does not fill a plane,
has a fractal dimension (Hausdorff-Besicovitch dimension, or just Hausdorff
dimension) greater than~1 but
less than~2; a fractal surface (e.g.~a fracture surface) will have a fractal
dimension between~2 and~3, and so on. \cite{federJ:fra-1988} showed, by
analyzing wave data from  the Norwegian continental shelf, that
a time series of significant wave height can have a fractal behavior.
 More details on the definition and
calculation of fractal dimension are given in the next section.

If the graph of an environmental parameter $h(t)$ has fractal behavior, it
will have a fractal dimension between~1 and~2.  If this graph
crosses a horizontal straight line, it will intersect the
line infinitely many times.  If we consider a small interval around one of
the intersection points, there will always be more intersection points within
the interval, no matter how small the interval is.
The collection of points where $h(t)$ intersects the
line $h=h_0$ is called the {\em level set} of $h(t)$ at $h_0$, and has a
fractal
dimension between~0 and~1.  A point has dimension zero, as has a collection
of points which are all more than a certain distance apart, but this particular
collection of points will be so numerous and irregularly distributed that
it will have a fractional dimension strictly greater than zero.

Since
the number of intervals satisfying $h<h_0$ is infinite,
any estimate of $\overline\tau$ using values of $h(t)$ sampled at successive
recording intervals $\Delta t$ will become smaller and smaller as $\Delta t
\to 0$.  This type of behavior was found by
\cite{soukissianT:theochari:isope-2001-33} for
measured time series of significant
wave height and wave period. They derived formulas for mean duration of
the form $\overline\tau = a + b\log(\Delta
t/\delta_0)$, where $a$, $b$, and $\delta_0$ are constants.  However, this
logarithmic formula is not valid in the limit $\Delta t \to \rm0$, since
$\overline\tau$ cannot be negative.

\section*{\rm\normalsize DURATION DISTRIBUTION}
The only 
possible limits of $\overline\tau$ as $\Delta t \to \rm0$ are a finite 
positive value
or zero (we assume here that the measurements take place over a finite period
so that the maximum value of $\tau_i$ is finite). 
To make more specific predictions of the behavior of the duration statistics,
we shall now determine how they are related to the
Hausdorff-Besicovitch dimension \citep{hausdorff:ma-1919-157}
of the above-mentioned level sets of~$h(t)$.

Firstly, we define the
$d$-dimensional Hausdorff {\rm measure} $\Lambda^d(\mathcal{S})$ of a set
$\mathcal{S}$~\cite[e.g.][]{mandelbrotBB:fgn-1983,RevuzD:YorM:CMBM-1991}:
\begin{equation}
\Lambda^d(\mathcal{S}) = \lim_{\rho\to0}
\inf\left(\sum_i{\pi^{d/2}{\rho_i}^d\over\Gamma(1+{d\over2})
}\right),\qquad \rho_i \leq
\rho, \label{hausdorff-measure}
\end{equation}
where the infimum (greatest lower bound) is over all coverings of $\mathcal S$ 
by balls (solid spheres) of radius less than or equal to $\rho$.  (The quantity
${\pi^{d/2}{\rho_i}^d\over\Gamma(1+{d\over2})}$ 
is the $d$-dimensional generalization of the formula for the volume of a ball
of radius $\rho_i$.)
In other words, you cover the set with as few balls as possible no
greater than a certain size, add up their ``$d$-dimensional'' volumes,
and repeat the process with smaller and smaller balls.  If $d$ is
an integer, and $\mathcal S$ is a subset of $d$-dimensional Euclidean space,
then the Hausdorff measure is the $d$-dimensional Lebesgue measure,
equivalent to what we usually mean by length, area, volume, etc.

If $\Lambda^d(\mathcal{S}) = \rm0$ for $d>D$ and  $\Lambda^d(\mathcal{S}) =
\infty$ for $d<D$, then $D$ is said to be the Hausdorff dimension of
$\mathcal S$.  In
this case, $\Lambda^D(\mathcal{S})$ may be zero, finite, or infinite.
The level sets of $h(t)$ are sets of points embedded in a one-dimensional
space, so the ``balls'' in the definition of Hausdorff measure may be replaced
by intervals of length $s_i=2\rho_i$.

In the case of Brownian motion, where $h(t)$ can be thought of as
a one-dimensional random walk with infinitesimally small steps
\citep{levyP:psmb-1965}, the level sets 
have $D=\frac{1}{2}$.  Fractional Brownian motion $B_H(t)$, described by 
\cite{%
mandelbrotBB:vanNess:siamr-1968-422%
}, where $H$ is the Hurst exponent, $0<H<1$, and with
$H=\frac{1}{2}$ corresponding to ordinary Brownian
motion,
has level sets with $D=1-H$.  If $h(t)$ is a random function whose values at
different values of $t$ are completely uncorrelated, it will have level  
sets with $D=1$.

The simplest method of computing the Hausdorff dimension is by the
{\em box-counting method}, where we just divide
the time domain into fixed intervals of length $s$.  If $N$
such fixed intervals are required to cover the level set, then the
{box-counting
dimension\/} can be defined as minus the slope of the graph of
$\log N$ {\em v.}~$\log s$, were we can, for example, fit a straight line
to the graph of the data points.
(Strictly speaking, for the Hausdorff and
box-counting dimensions to be equal, we should compute the asymptotic slope for
small box sizes.)
We say that
\begin{equation}
N(s) = O(\phi(s)\cdot s^{-D}),
\label{boxcounting}
\end{equation}
where $\phi(s)$ is a function which varies more slowly than any power of
$s$. For simplicity we will henceforth neglect the presence of $\phi(s)$, and
use the notation $A \sim B$ to mean  $A = O(B)$.

The distribution of the lengths $\tau_i$ of intervals where the relation
$h<h_0$ is satisfied
depends on the distribution of points of the corresponding level set
$\mathcal{L}$,
since the end points of each of the intervals are also
points in $\mathcal{L}$.  If we cover
$\mathcal{L}$ with fixed intervals of length $s$, the total number of
fixed intervals required will also
be of the order of $N(\tau_i>s)$,
the number of intervals with duration greater than $s$.
But by the definition given above of
fractal dimension, the number of fixed intervals required is of order
$s^{-D}$.

Thus we have
\begin{equation}
N(\tau_i > s) \sim s^{-D}. \label{eq-Ntau-rho}
\end{equation}
For $0<D<1$ the mean duration is then given by
\begin{equation}
\overline\tau \sim \left(\int_{\tau_{\rm min}}^{\tau_{\rm max}}
s^{-D}\,ds \right)\bigg/
\left( \int_{\tau_{\rm min}}^{\tau_{\rm max}} s^{-D-1}
\,ds \right).
\end{equation}
The upper integration limit $\tau_{\rm max}$ can be taken to be
the total observation time
$T$, and the lower limit can be taken to be the recording interval $\Delta t$.
We then have, for~$0<D<1$,
\begin{equation}
\begin{split}
\overline\tau &\sim\frac{T^{1-D}-{\Delta t}^{1-D} }{ {\Delta t}^{-D} - T^{-D}}
                  \cdot \frac{D}{ 1-D}
,\\
              &\sim T^{1-D} {\Delta t}^D   \label{eqmeandur}
 \cdot D/(1-D)\quad \text{since }
 T\gg\Delta t.
\end{split}
\end{equation}
For~$D=0$, the level set $\cal L$ will in general
have a finite number of points, so
$\overline\tau$ should become constant for sufficiently small $\Delta t$.

For~$D>0$, the mean duration tends to zero as $\Delta t \to 0$.  For marine
and offshore
engineering purposes this means that it is a parameter which is not
particularly informative if we record observations at frequent intervals, in
spite of the fact that we would expect the environmental data to be more useful
if sampled more often.  The reason for the poor behavior of $\overline\tau$
for
small values of $\Delta t$ is that it over-represents very short intervals
which contribute very little to the engineering work which could, for example,
be performed.
If we weight each interval by its length $\tau_i$, we obtain what we can
call the {\em useful mean duration\/}:
\begin{equation}
\overline\tau^u = \left(\sum\tau_i^2\right)\Big/
\left(\sum\tau_i\right)=
\frac{\int\!\! s^2\,dN(s)}{
              \int\!\! s\,dN(s)}
.  \label{useful-mean}
\end{equation}
Using \eqref{eq-Ntau-rho}, we obtain
$\overline\tau^u\sim \text{(const.)}\cdot T$
for~$0\leq D < 1$, independent of $\Delta t$.
The most useful way of presenting the distribution of duration of intervals
of length $\tau_i$
is to determine the fraction of the time
$F(\tau_i>s)$ occupied by the intervals
of more than a specified length {\frenchspacing
\cite[e.g.][]{grahamC:ce-1982-303}},
and this also requires weighting them according to their length:
\begin{equation}
F(\tau_i > \tau) = (1/T)\,E\left[\sum_{\tau_i > \tau}\tau_i\right]
= (1/T)\!\int^{\tau}_T\!\!s\,\,dN(\tau_i>s). 
\label{useful-fraction}
\end{equation}
(Note that the integration limits are reversed, because $N(\tau_i>s)$ is
a decreasing function of~$s$. The quantity $N(\tau_i \leq s)$ may be infinite.)

It should be noted that if $dh/dt$ in the Rice--Kac formula (Eq.~\ref{eq-rice})
is replaced by its discrete version $\Delta h/\Delta t =
 (h(t+\Delta t)-h(t))/\Delta t$, we obtain
the relation
\begin{equation}
\overline\tau \sim \left(E\left[\left|{\Delta h
\over\Delta t}\right|\right]\right)^{-1} \label{eq-discrete}
\end{equation}
\citep[e.g.][]{Jenkins:UMD-2002}.

\section*{\protect\normalsize\rm ANALYSIS OF FIELD MEASUREMENTS}
\subsection*{\bf\normalsize Feder's fractal wave analysis}
\cite{federJ:fra-1988} made a fractal analysis of 3 years of wave data
from Tromsø\-flaket on the Norwegian continental shelf.  After performing a
seasonal adjustment on the time series of the significant wave height, he
calculated a cumulative sum, effectively integrating the results with respect
to
time, to obtain a record which he found to have fractal behavior
with a Hurst exponent 
$H=0.92$, corresponding to a `level set' fractal dimension of
0.08.  The asymptotic fractal behavior was found for time scales
between 3~hours and 10~days.  However, the wave record itself is more
irregular, in fact very discontinuous: if the cumulative sum of the wave height
record has $D=0.08$, the wave height record itself, obtained by taking
differences, will have level sets which have a fractal dimension of
$\,\min(1+0.08, 1) =   1$.

Hence for any limiting wave height $h_0$ there may be periods of finite
duration where the wave height is crossing over and under $h_0$ 
at virtually every recording interval.  The asymptotic behavior
of $\overline\tau$ will thus be
$\overline\tau \sim \Delta t$ as $\Delta t \to 0$.

\begin{figure}[!h]
\begin{center}
\includegraphics[width=0.75\textwidth]{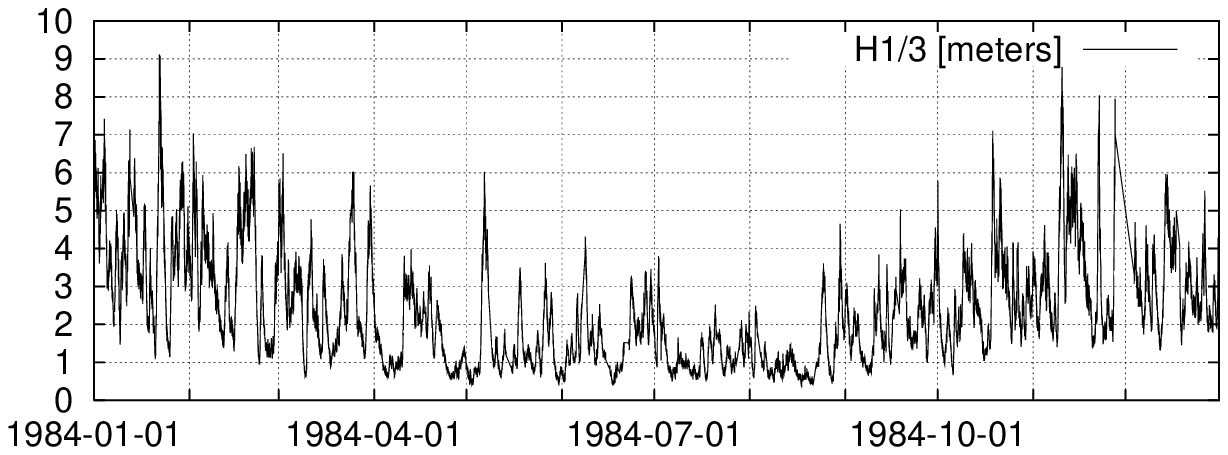}\\[1em]
\includegraphics[width=0.54\textwidth]{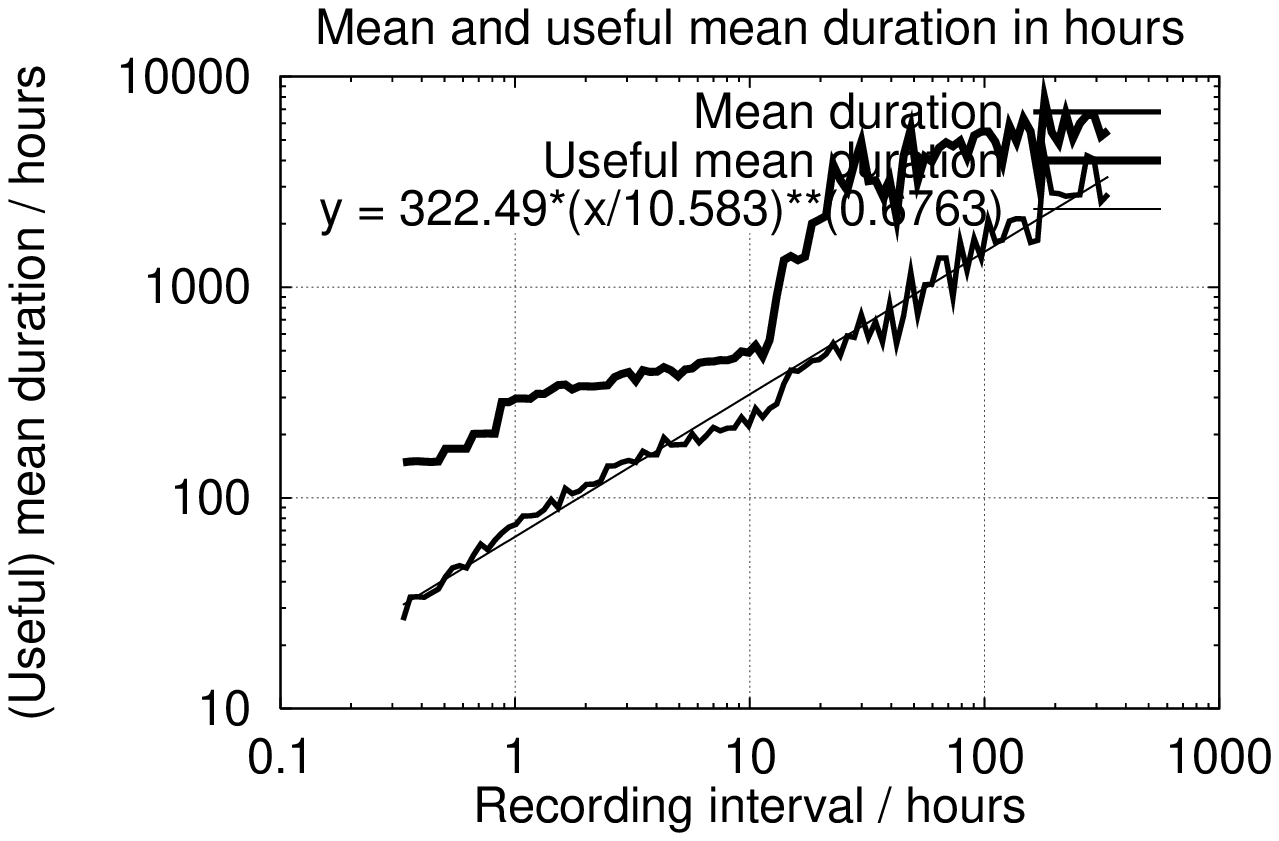}\\[1em]
\includegraphics[width=0.54\textwidth]{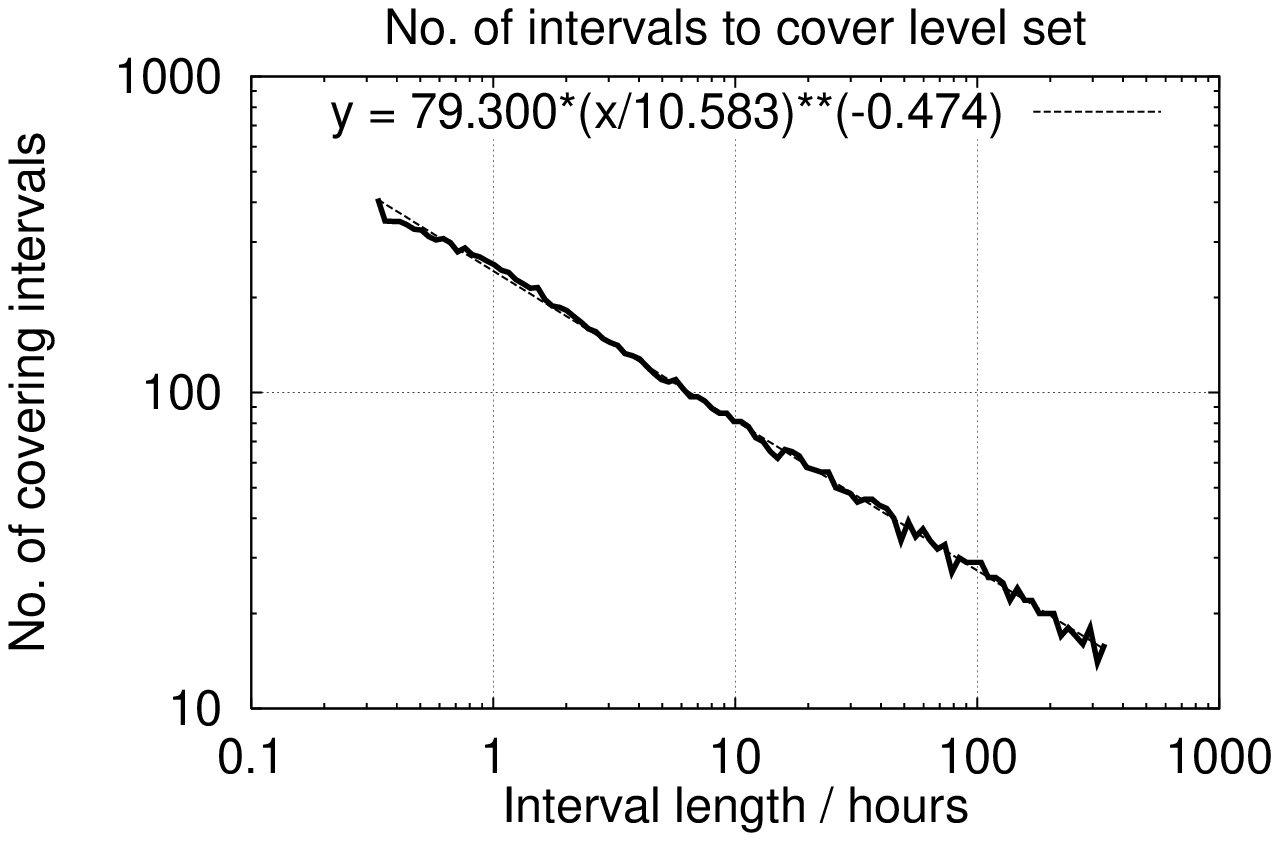}
\caption{\label{frigg-1984}\small
Top: Time series of significant wave height
($h=H_{1/3}$) from the Frigg gas field during 1984. Center: Mean duration and
useful mean duration for
intervals in the dataset with $h < \rm 5$\,m. Bottom: Number of fixed intervals
 necessary to cover the level set
of $h=\rm5$\,m, as a function of the interval length.
}
\end{center}
\end{figure}

\begin{figure}[!h]
\includegraphics[width=0.99\textwidth]{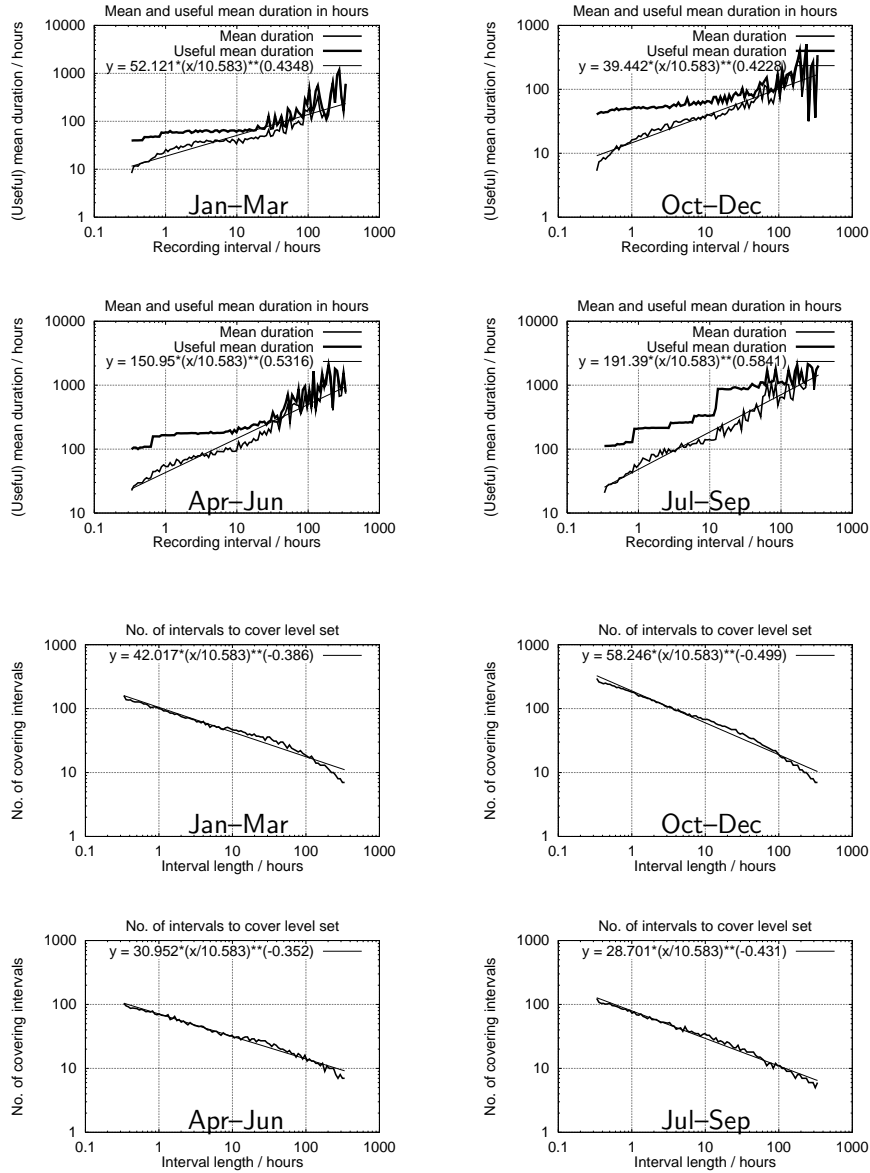}
\caption{\label{frigg-seasonal-meandur}\small
Upper graphs: Mean duration and useful mean duration for
intervals in the Frigg dataset with $h < 2.5\,$m, for successive
three-month periods. Lower graphs:
Number of fixed intervals necessary to cover the
corresponding $h=2.5\,$m level sets.
}
\end{figure}

\begin{figure}[!h]
\begin{center}
\includegraphics[width=0.75\textwidth]{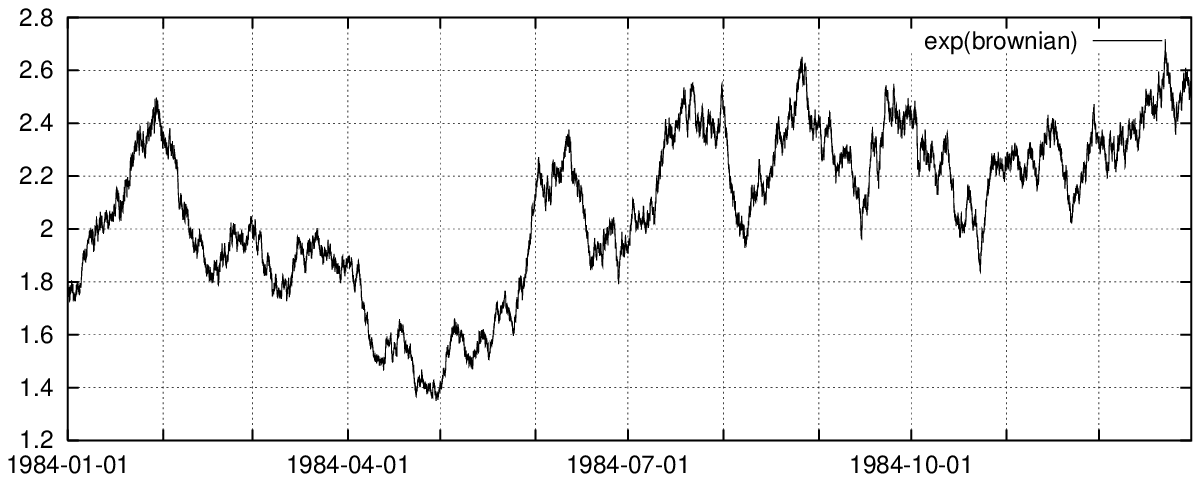}\\[1em]
\includegraphics[width=0.54\textwidth]{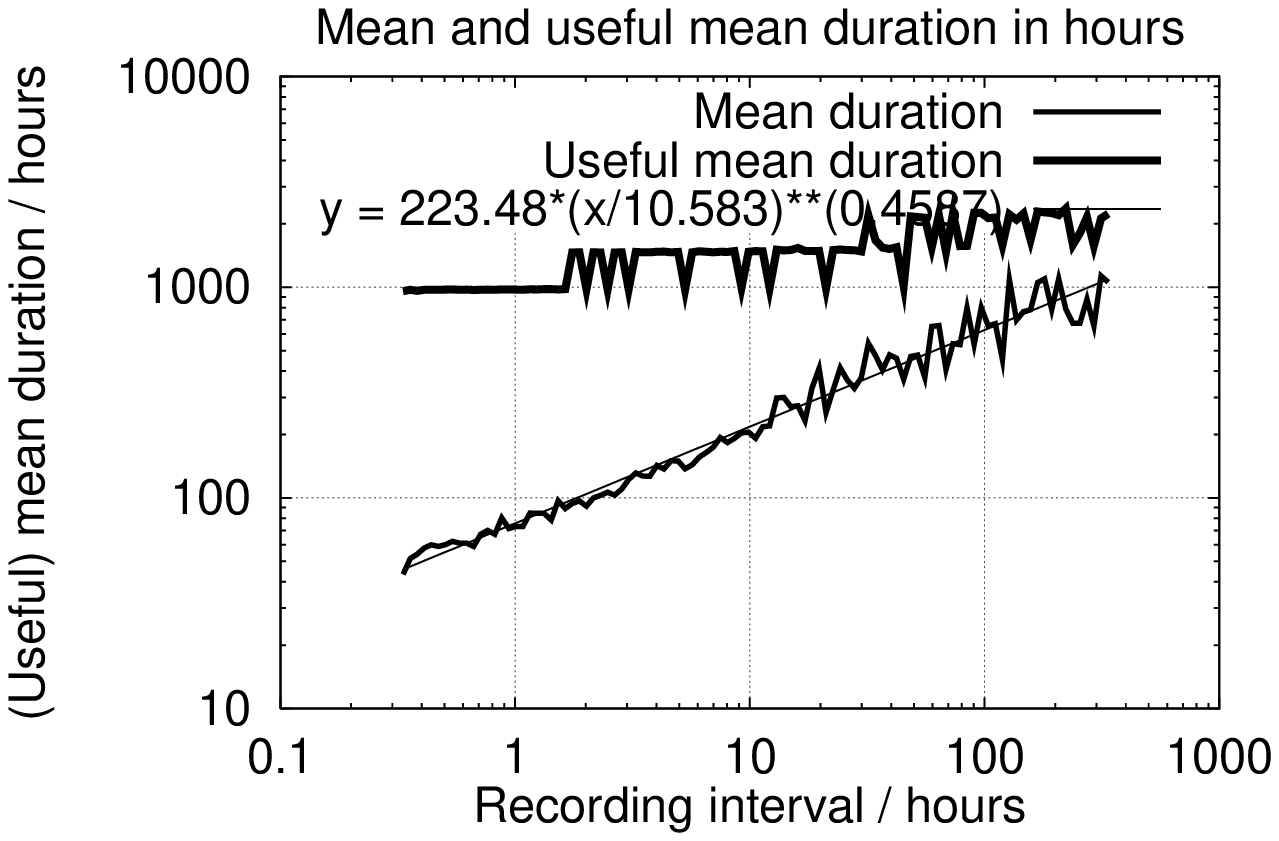}\\[1em]
\includegraphics[width=0.54\textwidth]{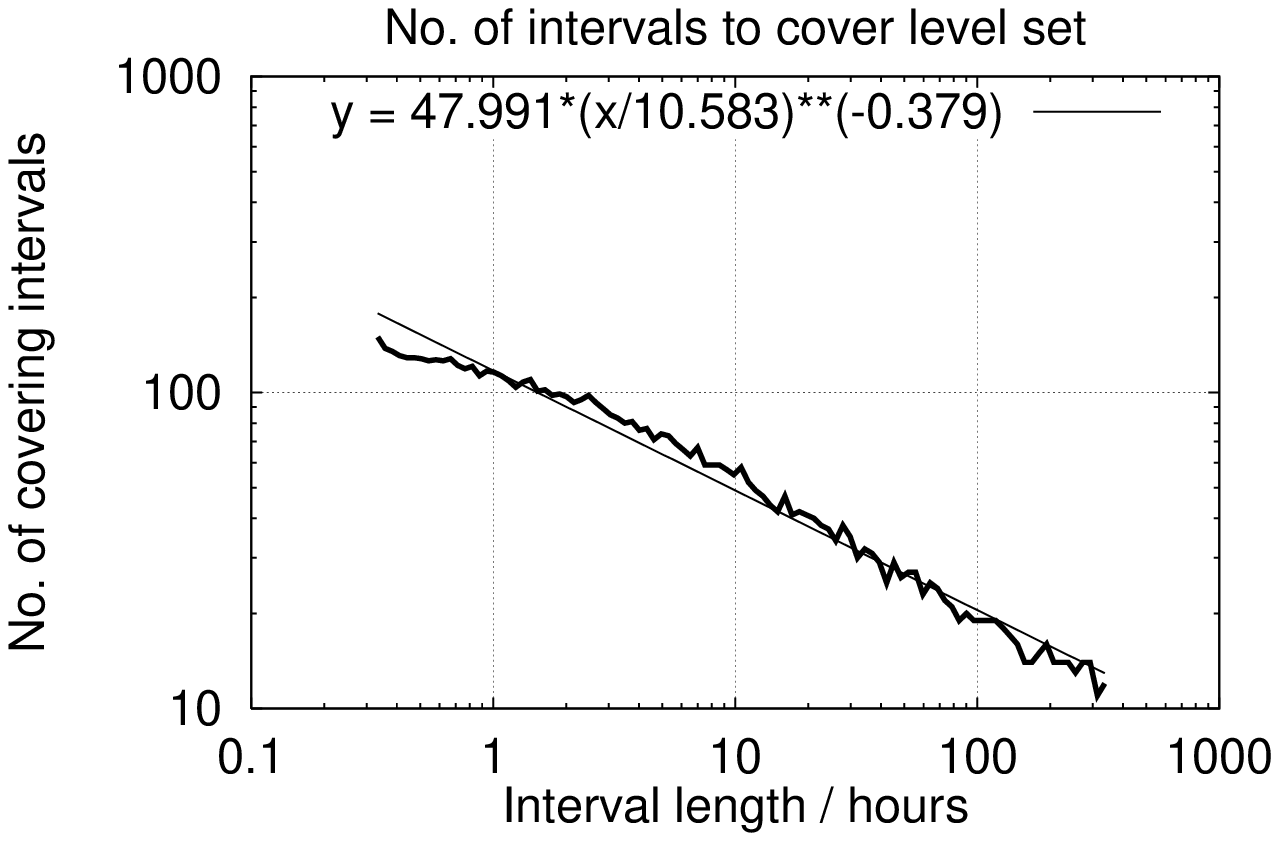}
\caption{\label{expbrown}\small
Top: Simulated time series (exponentiated Gaussian
random walk). Center: Mean duration and useful mean duration for
intervals in the simulated dataset with $h < \rm2.0$. Bottom: Number of fixed
intervals necessary to cover the level set
of $h=\rm2.0$, as a function of the interval length.
}
\end{center}
\end{figure}

\subsection*{\bf\normalsize Analysis of North Sea data}
Feder's results are, however,
based on measurements with the rather coarse time
interval of 3~hours, and his method of analysis, based on summing the
measurements, does not lead itself easily to the direct determination of either
the fractal dimension of the level sets, or of the calculation of the duration
statistics.  In the present study we analyze in a direct fashion a wave height
time series which has a finer, 20-minute time resolution.

The top frame of Fig.~\ref{frigg-1984} shows a time series of
significant wave height from the
Frigg gas field in the North Sea, for the year of 1984, based on an analysis of
20-minute records.  The time series was re-sampled at a range of different
recording intervals, from the initial 20 minutes up to 14 days (336 hours).
The mean duration and useful mean duration as a function of recording interval
are plotted in the center frame.  

The dependence of the mean duration on the recording interval is
fairly well approximated by a power-law relation, $\overline\tau \sim \Delta
t^{0.676}$.  From~\eqref{eqmeandur} we would therefore expect the
Hausdorff or the box-counting dimension of the level set of $h = h_0 =
5\,$m to
be $0.676$.  However, the straight-line fit by least-squares regression
in the lower frame of the figure shows,
according to \eqref{boxcounting}, that the box-counting dimension is
$D = 0.474$.  Similar values of the exponent of the power-law dependence of
mean duration on the recording interval, and box-counting dimension of the
corresponding level set, are obtained for other values of $h_0$.  The
alternative formula of Eq.~\ref{eq-discrete}, based on the Rice--Kac relation,
relating the mean duration to $|\Delta h/\Delta t|$, was also applied to the
same data set by \cite{Jenkins:UMD-2002}, and an intermediate exponent of~0.57
was obtained.

The useful mean duration $\overline\tau^u$
 also varies quite considerably with the recording
interval. There are a number of large jumps in the graph, which indicates
that what is happening is that
successive periods of $h<h_0$ are being amalgamated.  Nevertheless, the
variation of $\overline\tau^u$ with $\Delta t$ is less
rapid than that of the mean duration, and its value for small values of~$\Delta
t$ is about 150~hours (6 days), which may be a more reasonable estimate of a
typical ``$h<5\,$m'' period than the corresponding mean duration, which is
approximately 25~hours.

Figure~\ref{frigg-seasonal-meandur} shows the seasonal variation of the
corresponding statistics for the $h_0=2.5\,$m level, using four consecutive
three-month subsets of the data.   Both the mean duration and the useful mean
duration for $h<2.5\,$m are, as would be expected, greater in the summer 
than in the winter
months.  The discrepancy between the calculated box-counting dimension and the
exponent in the mean duration's power-law dependence are also present.
The graph of useful mean duration is noticeably flatter in the winter
months than in the summer months.

This discrepancy between the calculated box-counting dimension and the
exponent in the mean duration's power-law dependence also occurs for
simulated data, as shown in Fig.~\ref{expbrown}.
The simulated data are generated by firstly constructing an approximation to
Brownian motion $B(t)$ by generating a Gaussian random walk  with time steps of
20 minutes, and then converting the results to strictly positive values by
applying an exponential function, so we obtain a sample curve of $\exp(B(t))$.
The Hausdorff dimension
of the level sets of this graph
should be $D=0.5$, close to the calculated box-counting
dimension of the Frigg level set,  but the computed straight-line regression
value (for $h = 2.0$)  is here $0.379$, indicating that further
investigation is necessary
into what algorithm
is suitable for calculating Hausdorff dimension.
The exponent of the power-law dependence of the mean duration is 0.459, rather
greater than the calculated box-counting dimension.

\cite{Gorski:ph-2001-7933} has a useful
discussion of the limitations of the box-counting technique, and some
suggestions for improvement, for example by
only counting points separated by more than the box (fixed interval) size.
Strictly speaking, we should estimate the asymptotic slopes of the curves in
the log--log plots, by only considering recording intervals below a given
cut-off value.  However, given the limited amount of data available, it is
difficult to determine what a suitable cut-off value should be.

The graph of the useful mean duration for the simulated data is much flatter
than for the field data, so that for small values of~$\Delta t$
the ratio of $\overline\tau^u$ to $\overline\tau$ is much larger.  This may be
due to the fact that the graph of the simulated data has by definition zero
sampling variability, and so no ``random excursions'' will occur which may tend
to split up otherwise continuous periods with $h<h_0$.

\section*{\protect\normalsize\rm CONCLUSION}
The theoretical analysis presented here shows that
duration statistics of ocean waves and other
environmental parameters whose time dependence has an irregular,
noisy behavior, can be related to
 the fractal characteristics of the relevant time series.
In particular, the asymptotic behavior of the so-called
``mean duration'' of, for example, intervals where the wave height is below
a certain level, is in general strongly dependent on the recording interval
of the observations, and may have a power-law dependence
 with an exponent equal to
the Hausdorff-Besicovitch dimension or the box-counting dimension
of the associated level set.

Testing this hypothesis using a one-year data set of significant wave height
data from the Frigg gas field, sampled at 20-minute intervals, and
then re-sampled at different recording intervals up to a maximum of 14
days, indicates that there does appear to be power-law dependence of
the mean duration on the recording interval, but that the exponent may not
be the same as the box-counting dimension of the corresponding level set.
A similar discrepancy is found if simulated data are used (an exponential
function applied to a random walk), and more detailed studies are
required to uncover the reason for this.

Instead of using the mean duration to characterize the duration
statistics, an alternative {\em useful mean duration\/} is proposed,
weighting each interval with its actual duration. This is more
meaningful in terms of, for example, the amount of work which can be
performed during maritime engineering operations.  Estimates of this
parameter are found to be somewhat less sensitive to changes in the recording 
interval than the mean duration, and have an appropriate seasonal variation.
However, the useful mean duration still does vary quite considerably with
recording interval, and this is probably due to both the inevitable effect of
changes in the detectability of wave height variations, and the effect of
sampling variability.  The fact that the useful mean duration varies
much more slowly when simulated data are used does indicate that sampling 
variability is a contributory factor. However, further studies using 
other types of simulated time-series are necessary before any definite
conclusion can be drawn.

The theory of excursions of continuous-in-time stochastic processes, and the
intensity (probability per unit time) of transition from e.g.~the
state $h<h_0$ to the state
$h\geq h_0$ should also give useful results for
engineering purposes.  Suitable references to the theory include
\cite{RevuzD:YorM:CMBM-1991} and~\cite{MichnaZ:mmor-1999-335}.  The
relationship between excursion theory and the results described in this paper
is a topic for further investigation.

\section*{\protect\normalsize\rm ACKNOWLEDGEMENTS}
This work was partly
 supported by the Commission of the European Communities under
contract EVK3-CT-2000-00026 of the Fifth Framework Programme (MAXWAVE project).
The author is solely responsible for the work and it does not represent the
opinion of the Community.
The Community is not responsible for any use that
may be made of the data appearing herein.

The Frigg wave data were provided by ELF and all their partners in the Frigg
field.  Michel Olagnon of IFREMER very kindly provided the author with a
computer file of the processed data.  I thank anonymous referees for
constructive suggestions which improved the paper.



\end{document}